\documentstyle[prb,preprint,aps,tighten]{revtex} 
\input{psfig}

\begin{document}

\title{Photon-Assisted Transport Through Ultrasmall Quantum Dots:
Influence of Intradot Transitions}

\author{Ph. Brune$^a$, C. Bruder$^{a,b}$, and H. Schoeller$^a$}

\address{$^a$ Institut f\"ur Theoretische Festk\"orperphysik,
Universit\"at Karlsruhe, D-76128 Karlsruhe, Germany\\ $^b$
Physikalisches Institut, Universit\"at Bayreuth, D-95440 Bayreuth,
Germany}

\maketitle \date{\today}
\begin{abstract}
We study transport through one or two ultrasmall quantum dots with
discrete energy levels to which a time-dependent field is applied
(e.g., microwaves). The AC field causes photon-assisted tunneling and
also transitions between discrete energy levels of the dot. We treat
the problem by introducing a generalization of the rotating-wave
approximation to arbitrarily many levels. We calculate the dc-current
through one dot and find satisfactory agreement with recent
experiments by Oosterkamp {\it et al.}  [Phys. Rev. Lett. {\bf 78},
1536 (1997)].  In addition, we propose a novel electron pump
consisting of two serially coupled single-level quantum dots with a
time-dependent interdot barrier.
\end{abstract}
\section{Introduction}
Transport through small quantum dots has attracted considerable
interest over the last couple of years. These quantum dots, small
structures formed in a two-dimensional electron gas by applying
appropriate gate voltages, are characterized by small capacitances to
the substrate and to the leads connecting them to external voltage
sources. Hence there is a sizeable charging energy $E_C=e^2/(2C)$ that
has to be provided if electrons are to tunnel from the leads to the
dot. Transport is blocked at small voltages, a phenomenon dubbed the
{\it Coulomb blockade} since it is a direct consequence of the Coulomb
interaction and the geometry of the dot. Another aspect that comes up
for semiconductor quantum dots as opposed to small metallic islands is
their discrete single-particle spectrum caused by size quantization.

Many aspects of the Coulomb blockade are now well
understood. Recently, a new issue has come up, viz., time-dependent
transport through small quantum dots. High-frequency AC voltages can
be applied to mesoscopic structures (e.g., in the form of microwaves).
They lead to photon-assisted tunneling, i.e., electrons can overcome
the Coulomb blockade by absorbing photons from the external field.
This has become a very active area recently both experimentally
\cite{Kouwen a,Kouwen b,blick,kotthaus1,Oosterkamp} and theoretically
\cite{Buett,jauho,Bruder,Stoof,staffordwingreen,isawa}.

In this work, we will study transport through an ultrasmall quantum
dot with discrete energy levels to which a time-dependent field is
applied. The electron interaction in the dots is taken into account by
the Coulomb blockade model. The dots are weakly coupled to source and
drain reservoirs by tunnel junctions. Time-dependent gate voltages
lead to photon-assisted tunneling. In contrast to earlier theoretical
work \cite{Buett,jauho,Bruder,Stoof,staffordwingreen} we also take
into account transitions between discrete energy levels of the dot.
In addition, we propose a novel electron pump consisting of two
serially coupled single-level quantum dots strongly coupled by
time-dependent fields.

The paper is organized as follows: in Section \ref{model} we introduce
the Hamiltonian of a single interacting quantum dot with a
time-dependent field, connected by tunnel junctions to source and
drain reservoirs. We discuss the model and its solution by introducing
a generalized version of the rotating-wave approximation (RWA). In the
following section, we describe briefly the master equation technique
we use to calculate the transport current. In this paper, the
tunneling is always taken into account by performing a first-order
perturbation expansion in the tunneling matrix element. This is
equivalent to consider sequential tunneling, assuming the dot to be
weakly coupled to the reservoirs such that higher-order tunneling
processes can be neglected. In Section \ref{twolevel} we describe the
case of a dot with two discrete energy levels, which can be solved
analytically. Our results for the current are presented in Section
\ref{results} and compared with recent experiments \cite{Oosterkamp}.

In Section \ref{floquet} we present a double-dot electron pump, which
uses a time-dependent interdot barrier as the pumping mechanism. We
use the Floquet-matrix technique to find a numerical solution to the
problem, valid even in situations in which the RWA is not applicable.

\section{Model}
\label{model}
As a model for an interacting quantum dot in a time-dependent periodic
field coupled to two reservoirs by tunnel junctions we will use the
time-dependent Hamiltonian
\mbox{$H(t)=H_{res}(t)+H_{dot}(t)+H_{tun}$\cite{Bruder}}. Here
\begin{equation} 
H_{res}(t)=\sum_{k,\alpha,\sigma}\epsilon^{\phantom{\dagger}}_{k\alpha\sigma}
(t)\, c^\dagger_{k\alpha\sigma} c^{\phantom{\dagger}}_{k\alpha\sigma}
\end{equation}
describes noninteracting electrons in the reservoirs
$\{\alpha\}=\{L,R\}$,
$c^\dagger_{k\alpha\sigma}/c^{\phantom{\dagger}}_{k\alpha\sigma}$ are
the creation/annihilation operators of an electron with momentum $k$
and spin $\sigma$ in the reservoir $\alpha$.  The energies
$\epsilon^{\phantom{\dagger}}_{k\alpha\sigma}(t)=
\epsilon^0_{k\alpha\sigma}+\Delta_\alpha \cos(\omega t)$ include a
time-dependent shift of the Fermi energy of the electrons in the leads
due to the applied periodic field, and $\Delta_\alpha$ denotes its
coupling strength to the reservoir $\alpha$.  The Hamiltonian for the
interacting electrons in the dot is given by
\begin{equation} 
H_{dot}(t)=\sum_{l,\sigma}\epsilon^{\phantom{\dagger}}_{l\sigma}(t) \,
 d^\dagger_{l\sigma} d^{\phantom{\dagger}}_{l\sigma} +
 \sum^{m<l}_{l,m,\sigma}w^{\phantom{\dagger}}_{lm}(t) \,
 d^\dagger_{l\sigma} d^{\phantom{\dagger}}_{m\sigma}+
 h.c.+H_{ch}(N_{dot})\; ,
\label{Hdot}
\end{equation}
where $d^\dagger_{l\sigma}/d^{\phantom{\dagger}}_{l\sigma}$
create/annihilate electrons with spin $\sigma$ occupying level $l$ of
the discrete, equidistant energy levels with level spacing
$\Delta\epsilon$ in the dot ($l=1,2,... ,N$ for a quantum dot with $N$
levels).  The energy of level $l$ is given by
$\epsilon^{\phantom{\dagger}}_{l\sigma}(t)=\epsilon^0_{l\sigma}+
\Delta_D \cos(\omega t)$, where the time dependence is taken into
account by a periodic shift of the level. The coupling strength of the
field to the dot is given by $\Delta_D$. The time-dependent transition
matrix elements $w^{\phantom{\dagger}}_{lm}(t)=\overline{\Delta}_{lm}
\cos(\omega t)$ describe transitions from level $l$ to level $m$,
i.e., transitions that do not change the number of electrons in the
dot.  The Coulomb interaction between electrons in the dot is taken
into account by the Coulomb-blockade model
\begin{equation} 
H_{ch}(N_{dot})=E_C N_{dot}^2\; .
\end{equation}
Here, $N_{dot}=\sum_{l,\sigma} d^\dagger_{l\sigma}
d^{\phantom{\dagger}}_{l\sigma}$ is the particle number in the dot,
$E_C= e^2/2C$ is the charging energy with $C = C_L+C_R+C_g$. Note that
we have already taken into account the time-dependent part $2
N_{dot}\,n_0(t)$ of the original
$H^{orig}_{ch}(N_{dot},t)=E_C[N_{dot}+n_0(t)]^2$ by the energies
$\epsilon^{\phantom{\dagger}}_{l\sigma}(t)$ \cite{Bruder}. There $e
n_0(t)=C_L V_L(t)+C_R V_R(t)+C_g V_g(t)$ is related to the
polarization charges produced by the time-dependent voltages of the
left and right reservoirs $e V_{L/R}(t)=\mu_{L/R}+\Delta_{L/R}
\cos(\omega t)$ as well as the time-dependent gate voltage $e
V_g(t)=\mu_g+ \Delta_g \cos(\omega t)$ applied to the quantum dot by
the capacitance $C_g$.  The tunneling part is given by
\begin{equation} 
H_{tun}=\sum_{k,l,\alpha,\sigma}T^\alpha_{kl} \,
 c^\dagger_{k\alpha\sigma} d^{\phantom{\dagger}}_{l\sigma}+h.c.\; ,
\end{equation}
where $T^\alpha_{kl}$ denotes the tunneling matrix element.

The time-dependent Schr\"odinger equation ($\hbar=1$)
\begin{equation}
i{\partial \over \partial t}|\psi\rangle = H_0|\psi\rangle \qquad
H_0(t)=H_{res}(t)+H_{dot}(t)
\label{SchrGl}
\end{equation}
cannot be solved in a closed form due to the time-dependent
off-diagonal matrix elements $w_{lm}(t)$ in Eq. (\ref{Hdot}). For
\begin{equation}
\Delta\epsilon \approx \omega \; ,
\label{cond}
\end{equation}
one can approximate $w_{lm}(t)$ by
\begin{equation}
w^{\phantom{\dagger}}_{lm}(t)=\overline{\Delta}_{lm} \cos(\omega t)
={\overline{\Delta}_{lm}\over 2}(e^{i\omega t}+e^{-i\omega t}) \;\;
\longrightarrow \;\; w^{\phantom{\dagger}}_{lm}(t)={\Delta_{lm}\over
2} e^{-i\omega (l-m)t} \; .
\end{equation}
This is equivalent to omitting rapidly oscillating terms of frequency
$\omega+\Delta\epsilon$, which is much larger than
$\omega-\Delta\epsilon$ as long as (\ref{cond}) is fulfilled.

This can be understood as a generalization of the rotating-wave
approximation, which is well-known in the theory of time-dependent
two-level systems (e.g., in nuclear magnetic resonance
\cite{BlochSiegert} or quantum optics \cite{JaynCumm}). This
generalization can be applied to systems with {\it arbitrarily many}
levels.  It makes it possible to perform a time-dependent unitary
transformation $U(t)=U_d \, V(t)$ in Eq. (\ref{SchrGl}) which removes
the time dependence from the non-diagonal matrix elements of $H_{dot}$
[$V(t)$] and diagonalizes it afterwards [$U_d$]. Defining
$|\widetilde{\psi}\rangle=U(t)|\psi\rangle$, the Schr\"odinger
equation becomes
\begin{equation}
i{\partial \over \partial t}|\widetilde{\psi}\rangle =
\widetilde{H}_0(t)|\widetilde{\psi}\rangle \qquad
\widetilde{H}_0(t)=H_{res}+\widetilde{H}_{dot}(t)
\end{equation}
with
\begin{equation}
\widetilde{H}_{dot} =UH_{dot}U^\dagger-iU({\partial \over \partial
t}U^\dagger) \; ,
\label{Hdiag}
\end{equation}
where $UH_{dot}U^\dagger=UH_{dot}'U^\dagger+H_{ch}(N_{dot})$.  The
charging part of the Hamiltonian, $H_{ch}$, stays invariant under
unitary transformations, since it depends only on the particle number
on the dot.  $H_{dot}'$ is given by the following expression in matrix
notation
\begin{equation}
H_{dot}'(t)= \left(
\begin{array}{cc} 
H^{\uparrow}_{dot}(t) & 0 \\ 0 & H^{\downarrow}_{dot}(t)
\end{array}
\right)\; ,
\end{equation}
where the submatrices $H^{\sigma}_{dot}$ with spin index
$\sigma=\{\uparrow,\downarrow\}$ are given by
\begin{equation}
H^{\sigma}_{dot}(t)={1\over 2} \left(
\begin{array}{ccccc} 
2\epsilon^{\phantom{\dagger}}_{1\sigma}(t) & \Delta^*_1 e^{i\omega t}
& \Delta^*_2 e^{i2\omega} & \cdots & \Delta^*_{N-1} e^{i(N-1)\omega}\\
\Delta_1 e^{-i\omega t} & 2\epsilon^{\phantom{\dagger}}_{2\sigma}(t) &
\Delta^*_1 e^{i\omega t} & \cdots & \Delta^*_{N-2} e^{i(N-2)\omega}\\
\Delta_2 e^{-i2\omega t} & \Delta_1 e^{-i\omega t} &
2\epsilon^{\phantom{\dagger}}_{3\sigma}(t) & \cdots & \Delta^*_{N-3}
e^{i(N-3)\omega}\\ \vdots & \vdots & \vdots & \ddots & \vdots \\
\Delta_{N-1} e^{-i(N-1)\omega t} & \Delta_{N-2} e^{-i(N-2)\omega t} &
\Delta_{N-3} e^{-i(N-3)\omega t} & \cdots &
2\epsilon^{\phantom{\dagger}}_{N\sigma}(t)
\end{array}
\right)\; .
\label{Hmatrix}
\end{equation}
Here we have assumed $\Delta_{lm}=\Delta_{|l-m|}$ for simplicity.
Then $V(t)$ is
\begin{equation}
V(t)= \left(
\begin{array}{cc} 
V^{\uparrow}(t) & 0 \\ 0 & V^{\downarrow}(t)
\end{array}
\right)
\end{equation}
\begin{equation}
V^{\sigma}(t)=\left(\begin{array}{ccccc} e^{-i\omega t/2} & 0 & 0 &
\cdots & 0 \\ 0 & e^{i\omega t/2} & 0 & \cdots & 0 \\ 0 & 0 &
e^{i3\omega t/2} & \cdots & 0 \\ \vdots & \vdots & \vdots & \ddots &
\vdots \\ 0 & 0 & 0 & \cdots & e^{i(2N-3)\omega t/2}
\end{array}
\right)\;.
\end{equation}
The eigenenergies $\widetilde{\epsilon}^0_{j\sigma}$ ($j=1,2,... ,N$)
of $\widetilde{H}_{dot}(t)$ are then obtained by numerical
diagonalization.  The transformed tunneling part of the Hamiltonian is
given by
\begin{equation}
\widetilde{H}_{tun}(t)=\sum_{k,j,\alpha,\sigma}\widetilde{T}^\alpha_{kj}(t)\,
c^\dagger_{k\alpha\sigma}
\widetilde{d}^{\phantom{\dagger}}_{j\sigma}+h.c.  \;,
\label{Htun}
\end{equation}
where
$\widetilde{d}^\dagger_{j\sigma}/\widetilde{d}^{\phantom{\dagger}}_{j\sigma}$
are the creation/annihilation operators of an electron with spin
$\sigma$ that occupies the level with energy
$\widetilde{\epsilon}^0_{j\sigma}$, and
\begin{equation}
\widetilde{T}^\alpha_{kj}(t)=\sum_l {(U_d)}^*_{jl} \, T^\alpha_{kl} \,
e^{-i(2l-3)\omega t/2}\; .
\end{equation}
In addition, for a dot with only two levels the diagonalization can
also be performed analytically, providing further insight in the
underlying physics (see below in Section \ref{twolevel}).

\section{The master-equation approach}
$\widetilde{H}_{dot}(t)$ generates the time-evolution operator
$U_0(t,t_0)=\exp(-i\int_{t_0}^t d\tau \widetilde{H}_0(\tau))$, which
is needed to calculate the tunneling Hamiltonian
$\widetilde{H}^{I\alpha}_{tun}(t)=
U_0(t_0,t)\widetilde{H}^\alpha_{tun}(t)U_0(t,t_0)$ in the interaction
representation.

The von Neumann-equation $i \dot\varrho(t)=[\widetilde{H}_0(t)
+\widetilde{H}_{tun}(t),\varrho(t)]$ that describes the time evolution
of the density matrix is also transformed to the interaction
representation and $\varrho$ is expanded to first order in the
tunneling rate.  This leads to a master equation for the occupation
probabilities \cite{Bruder} $P_s(t)=\langle s|\varrho(t)|s \rangle$ of
the occupation number states $|s \rangle$. The states $|s \rangle$
represent the occupation numbers of the energy levels of the
diagonalized system. In the time-averaged dc-case the master equation
can be written as a system of coupled linear equations
\begin{equation}
\sum_{s'}[\Gamma_{s' \rightarrow s}-\delta_{ss'}(\sum_{s''} \Gamma_{s'
\rightarrow s''})] P_{s'} = 0\; ,
\end{equation} 
which can be solved approximately by a suitable truncation. The rate
$\Gamma_{s' \rightarrow s}$ for a transition from state $|s' \rangle$
to $|s \rangle$ can be expressed as $\Gamma_{s' \rightarrow
s}=\Gamma^L_{s' \rightarrow s}+ \Gamma^R_{s' \rightarrow s}$, where
\begin{equation}
\Gamma^\alpha_{s' \rightarrow s}={\omega\over \pi}
\int^{2\pi/\omega}_0 dt \int^\infty_0 d\tau \; Re\{\,\langle
s|\widetilde{H}^{I\alpha}_{tun}(t)|s' \rangle \langle
s'|\widetilde{H}^{I\alpha}_{tun}(t-\tau)|s \rangle \,\}
\end{equation}
is the rate associated with tunneling processes from/to reservoir
$\alpha$.  $\widetilde{H}^{I\alpha}_{tun}(t)$ denotes the part of the
tunneling Hamiltonian (\ref{Htun}) in the interaction picture that
corresponds to that reservoir.

The dc-current through the junction connecting the dot with reservoir
$\alpha$ now can be expressed in terms of the occupation probabilities
and transition rates as well as $N_{dot}(s)$ (the particle number on
the dot while being in state $|s \rangle$) as
\begin{equation}
I^\alpha_{dc}=-e \sum_{s,s'} \Gamma^\alpha_{s' \rightarrow s}
[N_{dot}(s')-N_{dot}(s)] \, P_s'\; .
\end{equation}

By this means it is possible to numerically calculate the current as a
function of transport or gate voltage. However, to really understand
the resulting I-V-curves, it is helpful to have a closer look at the
analytically solvable case with only two energy levels in the dot.

\section{Two-level case}
\label{twolevel}
The matrix (\ref{Hmatrix}) reduces to
\begin{equation}
H^{\sigma}_{dot}={1 \over 2} \left(
\begin{array}{cc} 
2\epsilon^{\phantom{\dagger}}_{1\sigma}(t) & \Delta^* e^{i\omega t} \\
\Delta e^{-i\omega t} & 2\epsilon^{\phantom{\dagger}}_{2\sigma}(t)
\end{array}
\right)
\end{equation}
with
$\epsilon^{\phantom{\dagger}}_{l\sigma}(t)=\epsilon^0_{l\sigma}+\Delta_D
\cos(\omega t)$. The transformation that renders the non-diagonal
elements time-independent is given by
\begin{equation}
V^{\sigma}(t)= \left(\begin{array}{cc} e^{-i\omega t/2} & 0\\ 0 &
e^{i\omega t/2}
\end{array}
\right)\; .
\end{equation}
Applying $U(t)$ to the Schr\"odinger equation leads to
\begin{equation}
\widetilde{H}_{dot} = {1\over 2} U^{\phantom{\dagger}}_d \left(
\begin{array}{cccc} 
2\epsilon^{\phantom{\dagger}}_{1\uparrow}(t)+\omega & \Delta^* & 0 &
0\\ \Delta & 2\epsilon^{\phantom{\dagger}}_{2\uparrow}(t)-\omega & 0 &
0\\ 0 & 0 & 2\epsilon^{\phantom{\dagger}}_{1\downarrow}(t)+\omega &
\Delta^*\\ 0 & 0 & \Delta &
2\epsilon^{\phantom{\dagger}}_{2\downarrow}(t)-\omega\\
\end{array}
\right) U^\dagger_d\; .
\end{equation}
Calculating $U_d$ and the new energies
$\widetilde{\epsilon}_{j\sigma}(t)$ is now straightforward, they are
given by
\begin{equation}
\widetilde{\epsilon}_{j\sigma}(t)={(\epsilon^0_{1\sigma}+\epsilon^0_{2\sigma})
\over 2}+{(-1)}^j {\Omega \over 2}+\Delta_D \cos(\omega t)\; ,
\end{equation}
where $\Omega=\sqrt{(\Delta \epsilon-\omega)^2+|\Delta|^2}$ is the
Rabi frequency and $\Delta \epsilon =
\epsilon^0_{2\sigma}-\epsilon^0_{1\sigma}$.  After calculating $U_d$
we can write down the creation/annihilation operators
$\widetilde{d}^\dagger_{j\sigma}/\widetilde{d}^{\phantom{\dagger}}_{j\sigma}$
for an electron that occupies energy level $j$ ($j=1,2$) as well as
the corresponding time-dependent tunneling matrix elements
$\widetilde{T}^\alpha_{kj}(t)$. Defining the quantity $\hat{\epsilon}=
\Delta \epsilon-\omega \pm \Omega$ for $\omega {\phantom{|}}^<_>
{\phantom{|}}\Delta \epsilon$, these matrix elements are
\begin{equation}
\widetilde{T}^\alpha_{k1}(t)={1 \over \sqrt{\hat{\epsilon}^2+
|\Delta|^2}}\,[ T^\alpha_{k1} \, \hat{\epsilon} \, e^{i\omega t/2} -
T^\alpha_{k2} \, \Delta \, e^{-i\omega t/2}]
\end{equation} 
\begin{equation}
\widetilde{T}^\alpha_{k2}(t)={1 \over \sqrt{\hat{\epsilon}^2+
|\Delta|^2}} \,[ T^\alpha_{k2} \, \hat{\epsilon} \, e^{-i\omega t/2} +
T^\alpha_{k1} \, \Delta^* \, e^{i\omega t/2}]\; .
\end{equation}

\begin{figure}
\centerline{\psfig{file=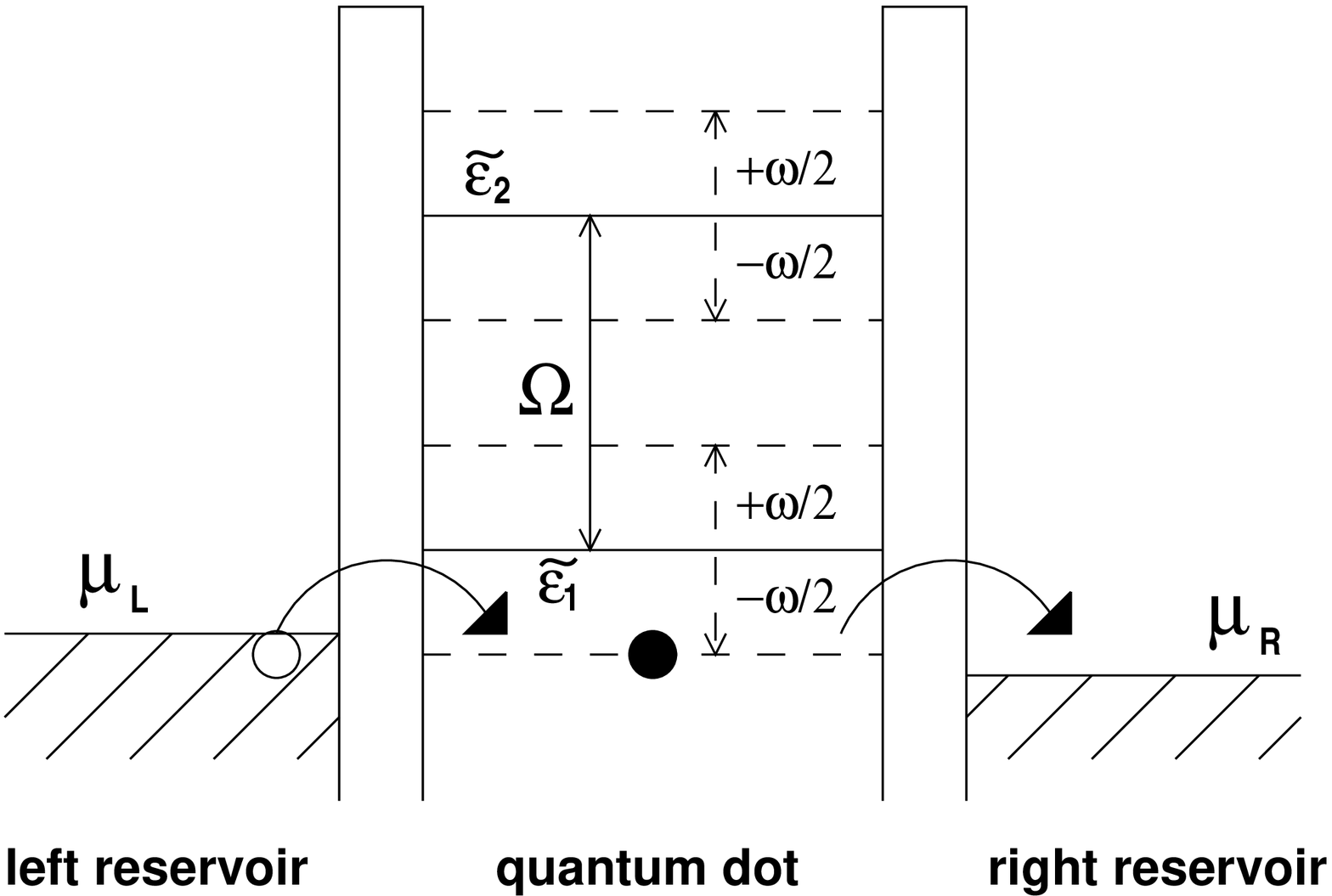,height=65mm,silent=}}
\vspace{5mm}
\caption{Energy landscape of a quantum dot with two non-degenerate levels
and inner transitions induced by a time-dependent field of frequency $\omega$.
The effective energy levels $\widetilde{\epsilon}_j$ are shown (solid lines).
Note that electron transport occurs, when the chemical 
potentials $\mu_{L/R}$ match one of the quasi-levels (dashed lines) 
shifted from the energies $\widetilde{\epsilon}_j$ by 
$\pm \omega/2$ due to the time-dependent tunneling matrix elements 
$\widetilde{T}^\alpha_{kj}(t)$.}
\label{fig:dot_two}
\end{figure}

Writing down the tunneling Hamiltonian with these matrix elements and
creation/\-annihilation operators and inserting them into the master
equation then permits us to calculate the current in a straightforward
way. It turns out that there are two possible ways for an electron to
tunnel through the dot, corresponding to the two terms with $e^{\pm i
\omega t/2}$ in the tunneling matrix elements
(Fig.~\ref{fig:dot_two}). The transport peaks present in the absence
of a time-dependent field are split in a two-peak group (with peaks at
distance $\omega$) that also shifts its position due to the
$\Delta$-dependence of the energies
$\widetilde{\epsilon}_{j\sigma}(t)$.

\section{Results}
\label{results}
For a dot with two spin-degenerate levels four groups of current peaks
will appear in the I-V-curve, separated from each other by $2E_C$. If
the field-induced inner transitions between the levels are neglected
(all $\Delta_{|l-m|}=0$), there is one main peak per group accompanied
by Bessel-type sidebands at separations $n\omega$ (with $n=\pm 1,\pm
2,\pm 3,...$). These side peaks are due to photon-assisted tunneling
(PAT) \cite{Bruder}. The existence of these side peaks has been
recently verified experimentally \cite{Oosterkamp}.

If inner transitions are taken into account, our calculations show
that the single main peak will shift with increasing $\Delta_{|l-m|}$.
In addition, $N-1$ peaks will appear at distances $n\omega$ (with
$n=1,...,N-1$), see also Fig.~\ref{fig:dot_two}. This leads to a
totally different picture of the current-peak positions and heights,
the weight of the peaks shifts as well as their positions.

To visualize only the influence of inner transitions, we set
$\Delta_D=0$, i.e, there is no PAT, and plot the peak group
corresponding to a dot occupied by one additional electron as a
function of the increasing strength of the inner transitions
(increasing $\Delta_{|l-m|}$). Two cases, one with two and one with
three degenerate energy levels are shown (Fig.~\ref{fig:Peaks}). The
shift of the main peak and the appearance of the one (two) additional
peak(s) as described above is clearly visible.

\begin{figure}
\centerline{\psfig{file=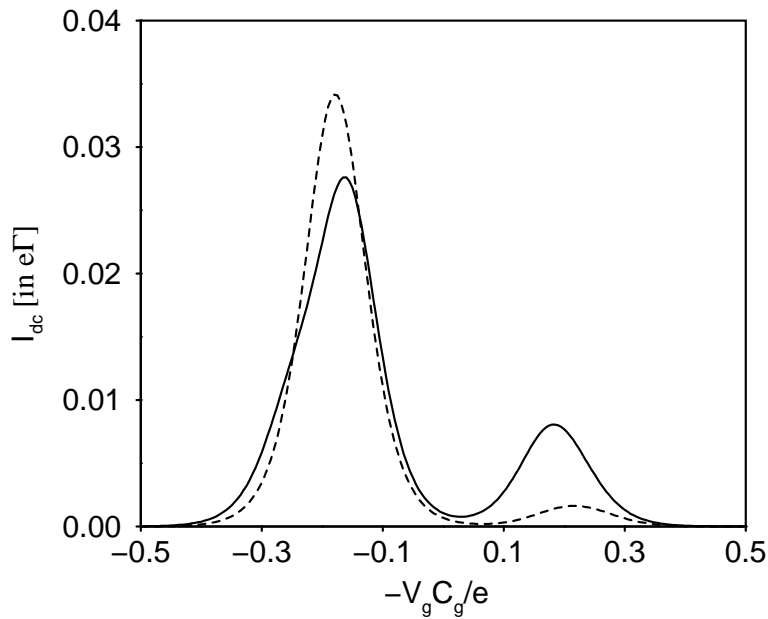,height=90mm}}
\caption{$I_{dc}$-$V_g$-curve for a dot with two degenerate levels.
The RWA model (solid curve) is compared with first-order perturbation 
theory for calculating the transition rates between the two energy levels
(dashed curve). 
In RWA the first side peak is significantly enhanced.
$\epsilon_1=-25\Gamma$, $\epsilon_2=25\Gamma$, $\omega=60\Gamma$, 
$T=5\Gamma$, $\Gamma_1=\Gamma_2=0.5\Gamma$, $\Delta_D=0$, $\Delta=10\Gamma$.}
\label{fig:vgl}
\end{figure}

Of course it is also possible to include the inner transitions in the
master equation in a perturbative way. To do this, transition rates
between the levels (similar to those describing the tunneling of
electrons to and from the dot) have to be calculated in first-order
perturbation theory.  Then the peaks do not shift with increasing
$\Delta_{|l-m|}$, because the master equation is written in the basis
of the unperturbed states. But also the effect on the peak heights is
significantly smaller than in our modified rotating-wave approximation
approach.  In Fig.~\ref{fig:vgl} this is illustrated by plotting again
the peak group for a dot with two levels and one extra electron.
Again we set $\Delta_D=0$, i.e., neglecting PAT. For a fixed,
relatively small value of $\Delta$, the peak shift is not significant.

\begin{figure}
\centerline{
\psfig{file=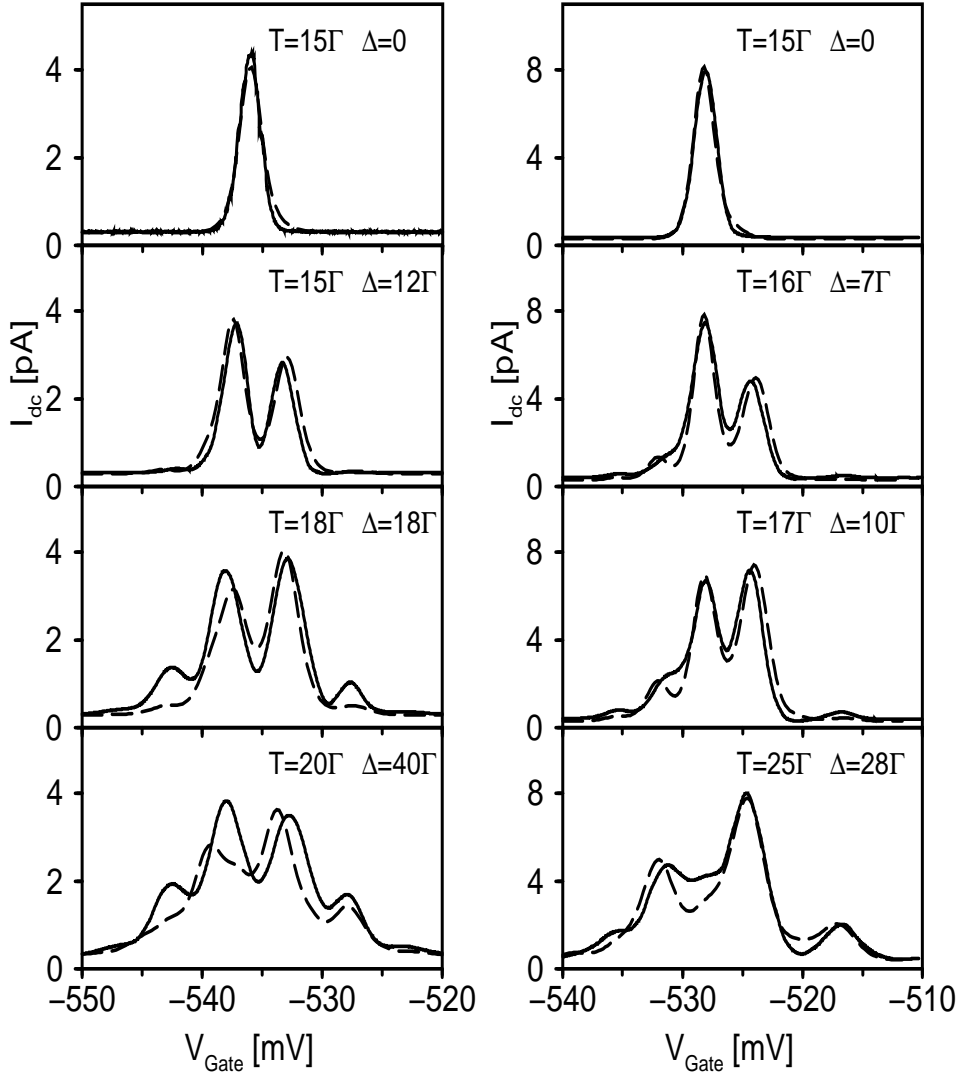,height=150mm,width=150mm,clip=,silent=}}
\caption{Measured $I_{dc}$-$V_g$-curves (solid lines) (Oosterkamp 
{\it et al.}\cite{Oosterkamp}) with applied
microwaves of frequencies $42\,$GHz$\,$(left column)
and $61.45\,$GHz$\,$(right column) versus our theoretical
results (dashed lines).
Left column: $\Gamma_1=0.2\Gamma$, $\Delta_D=1.875\Delta$.
Right column: $\Gamma_1=0.1\Gamma$, $\Delta_D=3.5\Delta$.
In both cases, $\Delta_L=-{1 \over 170}\Delta_D$, 
$\Delta_R=-{1 \over220}\Delta_D$.
Parameters determined by the experiment:
$\epsilon_1=-47.5\Gamma$, $\epsilon_2=47.5\Gamma$, $\omega=135\Gamma$ (left),
$\omega=200\Gamma$ (right), and $\Gamma=1.287\mu$eV.}
\label{fig:Messung}
\end{figure}

In recent transport measurements on ultrasmall quantum dots with
applied microwaves such a peak group was studied \cite{Oosterkamp}. A
dot with two levels contributing to transport was used. The first side
peak rises more strongly with increasing microwave power than expected
from the PAT model.  In Fig.~\ref{fig:Messung} we compare the measured
I-V-curves with our results. For small values of $\Delta$ our model is
in good agreement with the experiment for both microwave
frequencies. In particular it describes correctly the strong increase
of the right side peak. The numerical calculations have been done
using non-degenerate levels because in the experiments a strong
magnetic field ($B=0.91T$) (which was supposed to suppress plasmon
excitations) lifted the spin degeneracy.

\section{Double quantum-dot electron pump: Floquet-matrix approach}
\label{floquet}

The single quantum dot with two discrete energy levels connected by a
matrix element $w_{lm}(t)=\Delta_0+\Delta\cos(\omega t)$ can be mapped
to a double dot system where two dots with one energy level are
strongly coupled with each other. But in the formalism discussed above
it would not be possible in this case to choose the gate voltages and
the microwave frequency arbitrarily, due to the restrictions imposed
by the rotating-wave approximation. Also, the time-dependence of the
two gate voltages would have to be the same for both dots.

\begin{figure}
\vspace{10mm}
\centerline{\psfig{file=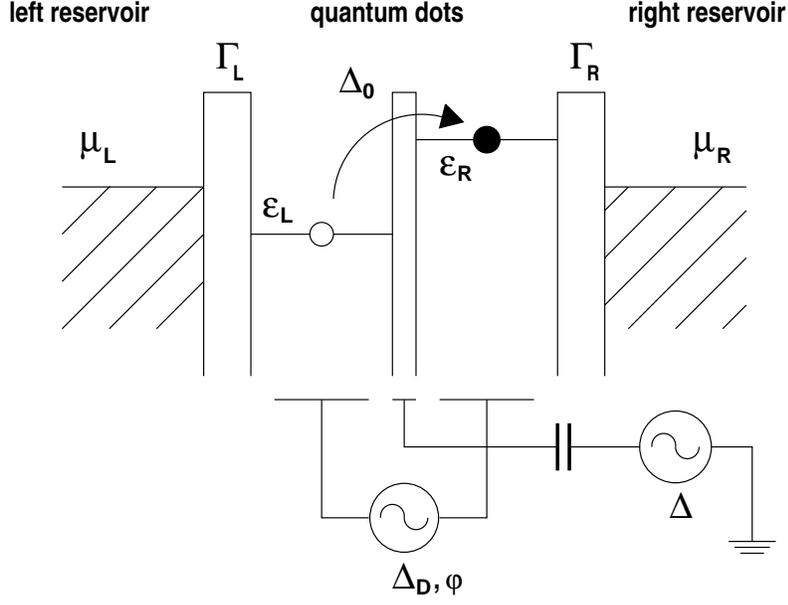,height=80mm,silent=}}
\vspace{5mm}
\caption{Energy landscape of two serially coupled quantum dots connected
by a weak time-dependent barrier (periodicity $\omega$). A
additional time dependence may also be applied to the gate electrodes 
of both dots. With $\mu_L=\mu_R$ and applied microwaves the system
works as an electron pump, pumping electrons ``uphill'' from left to right.}
\label{fig:ddot}
\end{figure}

\begin{figure}
\centerline{\psfig{file=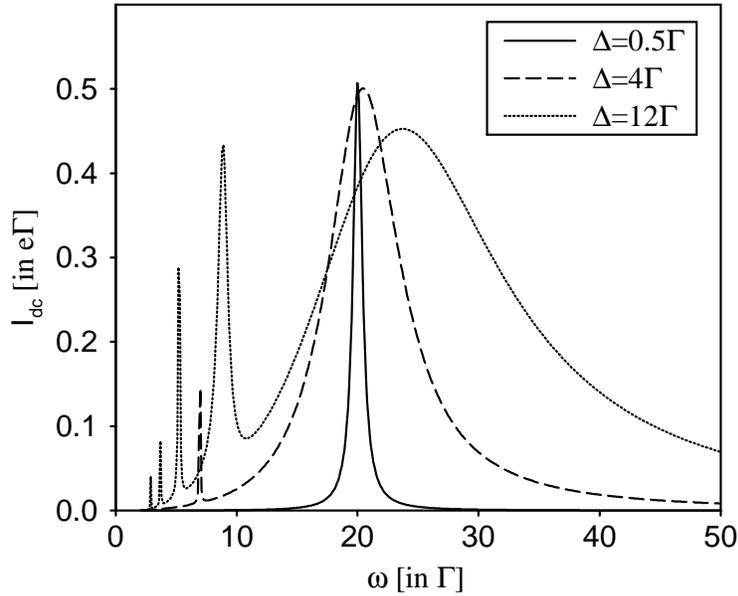,height=90mm,silent=}}
\caption{Current response of a double quantum-dot electron pump
versus frequency $\omega$ for a time-dependent barrier
separating the dots. $\Delta_0=0$, $\;\Delta_D=0$.}
\label{fig:pumpe}
\end{figure}

\begin{figure}
\centerline{\psfig{file=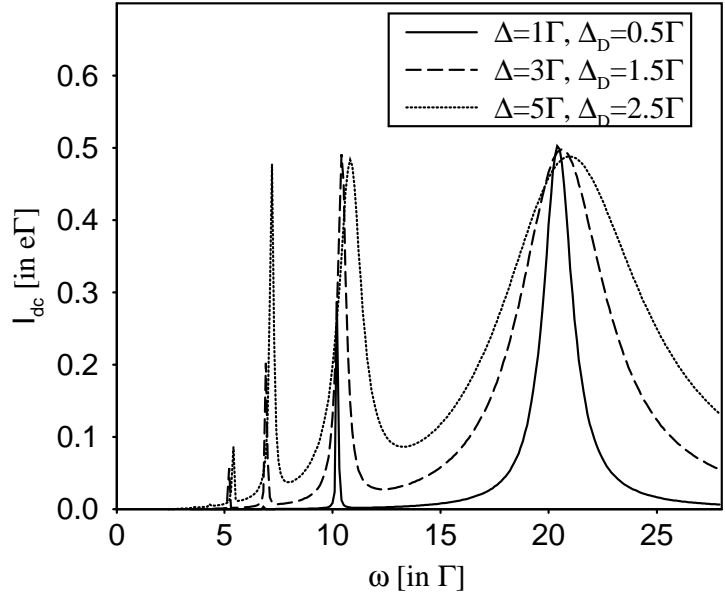,height=90mm,silent=}}
\caption{Same as Fig.~\protect\ref{fig:pumpe} 
with time-dependent gate voltages and assuming a phase difference of 
$\pi$ between the two gates. 
$\Delta_0=2\Gamma$.}
\label{fig:pumpe2}
\end{figure}

In this section we discuss a double dot system with time-dependent
gate voltages that differ by a relative phase $\varphi$. The dots are
strongly coupled by a time-dependent tunneling barrier .  Such a
system connected to reservoirs as shown in Fig. \ref{fig:ddot} can
work as a electron pump, resulting in a current even if no transport
voltage is applied.  We generalize the work of Stafford and
Wingreen\cite{staffordwingreen} to the case where the height of the
tunneling barrier also depends on time.  The time-dependent
Schr\"odinger equation is solved by the Floquet-matrix
approach\cite{Shirley}. In general, according to Floquet's theorem, a
differential equation with periodic coefficients like equation
(\ref{SchrGl}) has solutions of the form (for a dot with $N$ levels)
\cite{HoltHone}
\begin{equation}
|\psi_j(t)\rangle=e^{-i\tilde{\epsilon}_j t}\,
|\varphi_j(t)\rangle\;\;\;(j=1,...,N)\;,
\end{equation} 
which inserted in the Schr\"odinger equation (\ref{SchrGl}) result in
an eigenvalue problem for the states $|\varphi_j\rangle$
\begin{equation}
(H(t)-i{\partial \over \partial t})\,|\varphi_j\rangle =
\widetilde{\epsilon}_j |\varphi_j\rangle \;.
\end{equation} 
The states $|\varphi_j\rangle$ have the same periodicity as the dot
Hamiltonian, i.e.,
$|\varphi_j(t+2\pi/\omega)\rangle=|\varphi_j(t)\rangle$. Due to this
it is possible to expand $H_{dot}(t)$ and $|\varphi_j(t)\rangle$ in a
Fourier series
\begin{equation}
H_{dot}(t)=\sum_{n=-\infty}^\infty H_{dot}^{(n)} e^{in\omega t}
\label{H_FT}
\end{equation}
\begin{equation}
|\varphi_j(t)\rangle=\sum_{n=-\infty}^{\infty} e^{in\omega t}
|\varphi^{(n)}_j\rangle \;.
\end{equation}
In the basis $\{|\,l\,\rangle\}$ of the eigenstates of the diagonal
dot Hamiltonian $H^0_{dot}=\sum_{l,\sigma}\epsilon^0_l
d^{\dagger}_{l\sigma} d^{\phantom{\dagger}}_{l\sigma}$ with uncoupled
energy levels $\epsilon^0_l$ (i.e.,
$H^0_{dot}|\,l\,\rangle=\epsilon^0_l|\,l\,\rangle$), the solution
$|\psi_j(t)\rangle$ can be expressed as
\begin{equation}
|\psi_j(t)\rangle=\sum_{l=1}^N \langle\,l\,|\varphi_j(t)\rangle \,
e^{-i\tilde{\epsilon}_j t}\,|\,l\,\rangle =\sum_{l=1}^N
\sum_{n=-\infty}^{\infty} e^{-i\tilde{\epsilon}_j t} e^{in\omega t}
\,\varphi^{(n)}_{lj}\,|\,l\,\rangle \;.
\label{Psi_FT}
\end{equation} 
If we insert (\ref{H_FT}) and (\ref{Psi_FT}) in the time-dependent
Schr\"odinger equation (\ref{SchrGl}), multiply with the bra
$\langle\,i\,|$ from the left and define the matrix elements
$H^{(n)}_{il}=\langle\,i\,|H_{dot}^{(n)}|\, l\,\rangle$, we get an
infinite system of coupled linear equations describing the eigenvalue
problem for the quasi-energies $\tilde{\epsilon}_j$
\begin{equation}
\sum_{l=1}^N\sum_{k=-\infty}^{\infty}\,[H^{(n-k)}_{il}+n\omega\delta_{nk}
\delta_{il} ]\,\varphi^{(k)}_{lj} = \tilde{\epsilon}_j
\varphi^{(n)}_{ij} \;.
\end{equation}
If $H_{dot}(t)$ is given by (\ref{Hdot}) with
$w_{lm}(t)={\Delta_0}_{\,lm}+ \Delta_{lm}\cos(\omega t)$, this becomes
\begin{equation}
\sum_{l=1}^N\sum_{k=-\infty}^{\infty}\,[((\epsilon^0_l+n\omega)\delta_{il}
+{\Delta_0}_{il})\delta_{nk}+{\Delta_{il} \over 2}(\delta_{n,k+1}+
\delta_{n,k-1})]\, \varphi^{(k)}_{lj} = \tilde{\epsilon}_j
\varphi^{(n)}_{ij}
\label{ev}
\;.
\end{equation}
The quasi-energies $\tilde{\epsilon}_j$ and the eigenvector components
$\varphi^{(n)}_{ij}$ can be calculated numerically by truncating this
infinite system of coupled equations at a sufficiently large finite
$n$.  It is now possible to transform the tunneling Hamiltonian and
calculate the current by the master equation technique analogous to
the previous sections. The resulting dependence of the pumped current
on the frequency and the amplitude of the applied microwaves is
plotted in Fig. \ref{fig:pumpe} and Fig. \ref{fig:pumpe2}. Here and in
the rest of the paper we used
$E_{ch}=75\Gamma$,$\;\Gamma_L=\Gamma_R=\Gamma$.  In
Fig. \ref{fig:pumpe} and Fig. \ref{fig:pumpe2} we set
$\epsilon_1=-10\Gamma$,$\;\epsilon_2=10\Gamma$,$\;\mu_L=\mu_R=0$,
$\;T=5\Gamma$.

Figure \ref{fig:pumpe} illustrates the situation where the microwaves
couple only to the interdot barrier. There are current peaks if the
photon energy equals an integer fraction of the quasi-energy level
splitting. Because the quasi-energies themselves depend on the
amplitude $\Delta$ (analogous to the RWA calculations above), the
current peaks shift to higher photon energies with increasing
$\Delta$. This plot illustrates the effect created purely by a
time-dependent barrier.

However, in a real experiment the microwaves would also couple to the
gate electrodes and the interdot coupling would have a finite
time-independent part, i.e., $w_{lm}(t)=\Delta_0+ \Delta\cos(\omega
t)$. This is shown in Fig. \ref{fig:pumpe2}. It is clearly seen that
the overall peak height increases compared to Fig. \ref{fig:pumpe}.

\begin{figure}
\centerline{\psfig{file=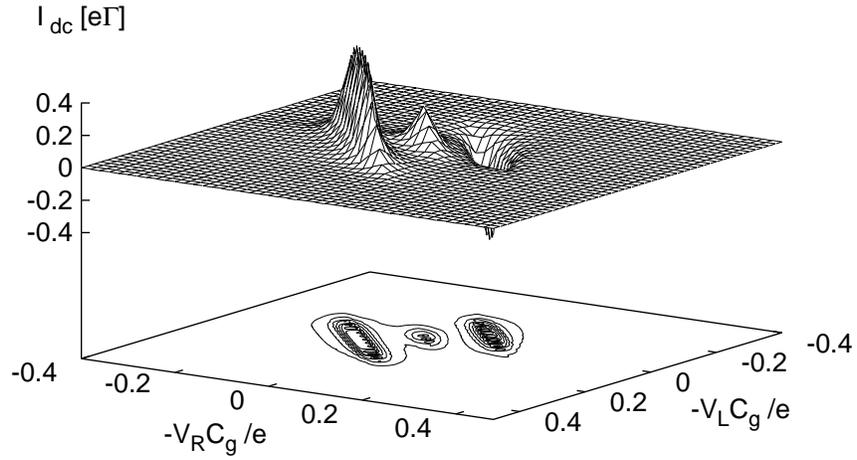,height=90mm,silent=}}
\caption{Dependence of the transport current on the positions of the 
energy levels, i.e., the gate voltages, for a time-dependent interdot barrier.
$\Delta_0=2\Gamma$,$\;\Delta=3\Gamma$,$\;\Delta_D=0$.}
\label{fig:trans1}
\end{figure}

\begin{figure}
\centerline{\psfig{file=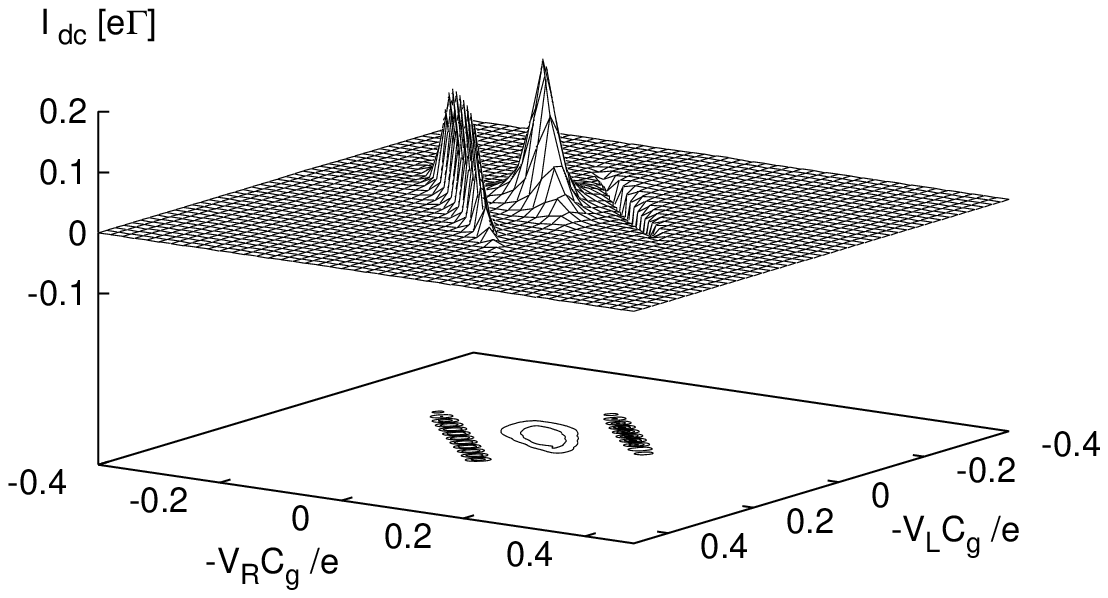,height=90mm,silent=}}
\caption{Same as Fig.~\protect\ref{fig:trans1}
for static interdot coupling ($\Delta=0$) and time-dependent gate voltages, 
assuming a phase difference of $\pi$ between the two gates.
$\Delta_0=2\Gamma$,$\;\Delta=0$,$\;\Delta_D=3\Gamma$.}
\label{fig:trans2}
\end{figure}

The behavior of the current through an interacting double quantum dot
for finite transport voltages and two independently varied gate
voltages is plotted in Fig. \ref{fig:trans1} and Fig. \ref{fig:trans2}
($\mu_L=2.5\Gamma$,$\;\mu_R=-2.5\Gamma$,$\;T=3\Gamma$ in both
figures).  Figure \ref{fig:trans1} illustrates the case where the
microwaves are coupled to the interdot barrier only, i.e., with a
time-dependent coupling matrix element. Instead, in
Fig. \ref{fig:trans2} the microwaves are coupled to the gate
electrodes, assuming a static interdot matrix element. As can be
clearly seen, the same value for the coupling matrix element in both
case leads to a significant increase of $I_{max}$.

\smallskip
In conclusion, we have calculated the photon-assisted transport
current through a single interacting quantum dot with an arbitrary
number of discrete energy levels. We have taken into account
field-induced inner transitions in a non-perturbative way by
generalizing the rotating-wave approximation to more then two energy
levels. We compare our results to recent experiments \cite{Oosterkamp}
and provide an explanation for the unexpected height of the first
side-band current peak.

We would like to thank J. K\"onig and G. Sch\"on for discussions and
suggestions and especially T.~H. Oosterkamp and L.~P. Kouwenhoven for
providing us with the experimental data. The support by the Deutsche
Forschungsgemeinschaft, through SFB 195, is gratefully acknowledged.
%

%
%

\begin{minipage}{\linewidth}
\begin{figure}
\centerline{\psfig{file=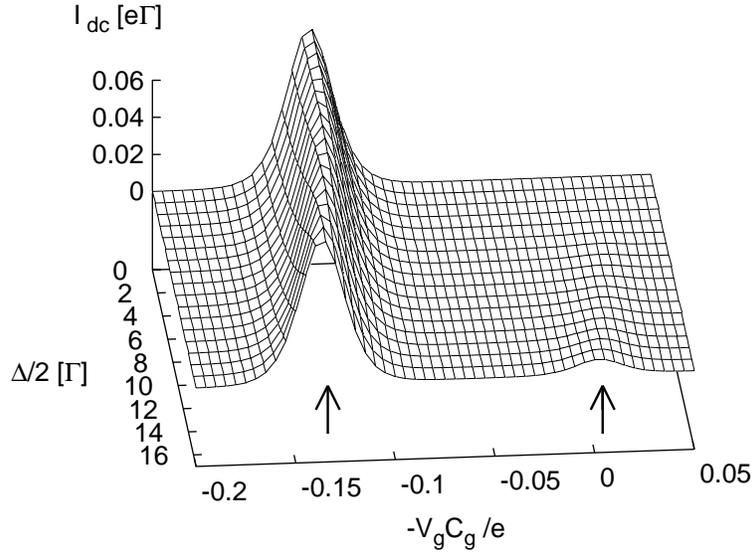,height=90mm,silent=}}
\centerline{\psfig{file=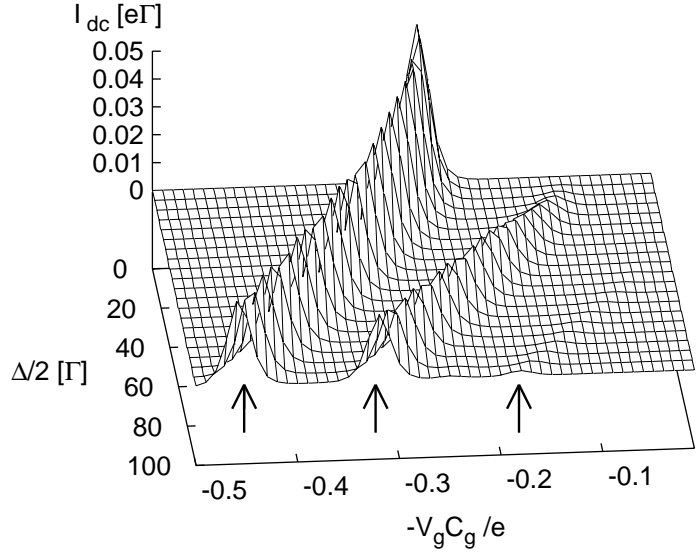,height=90mm,silent=}}
\caption{Evolution of the current-peak group corresponding to one 
electron occupying a dot with two (top) and three (bottom) spin-degenerate 
levels with increasing $\Delta$. PAT has been omitted. 
$\Delta_D=0$,$\;\Delta\epsilon=95\Gamma$,$\;\omega=55\Gamma$,$\;T=3\Gamma$.}
\label{fig:Peaks}
\end{figure}
\end{minipage}

\begin{references}

\bibitem{Oosterkamp} T.~H. Oosterkamp, L.~P. Kouwenhoven,
A.~E.~A. Koolen, N.~C. van der Vaart, and C.~J.~P.~M. Harmans,
Phys. Rev. Lett. {\bf 78}, 1536 (1997).

\bibitem{Kouwen a} L.~P. Kouwenhoven, S. Jauhar, K. McCormick,
D. Dixon, P.~L. McEuen, Y.~V.  Nazarov, N.~C. van der Vaart, and
C.~T. Foxon, Phys. Rev. B {\bf 50}, 2019 (1994).

\bibitem{Kouwen b} L.~P. Kouwenhoven, S. Jauhar, J. Orenstein,
P.~L. McEuen, Y. Nagamune, J. Motohisa, and H. Sakaki,
Phys. Rev. Lett. {\bf 73}, 3443 (1994).

\bibitem{blick} R.~H. Blick, R.~J. Haug, D.~W. van der Weide, K. von
Klitzing, and K. Ebert, Appl. Phys. Lett. {\bf 67}, 3924 (1995);
R.~H. Blick, R.~J. Haug, K. von Klitzing, and K. Ebert, Surf. Science
{\bf 361} 595 (1996).

\bibitem{kotthaus1} K. Fujii, W. Goedel, D.~A. Wharam, S. Manus,
J.~P. Kotthaus, G. B\"ohm, W. Klein, G. Tr\"ankle, and G. Weimann,
Physica B {\bf 227}, 98 (1996).

\bibitem{Buett} M. B\"uttiker, A. Pr\^etre, and H. Thomas,
Phys. Rev. Lett. {\bf 70}, 4114 (1993).

\bibitem{jauho} N.~S. Wingreen, A. Jauho, and Y. Meir, Phys. Rev. B
{\bf 48}, 8487 (1993); A.Jauho, N.~S. Wingreen, and Y. Meir,
Phys. Rev. B {\bf 50}, 5528 (1994).

\bibitem{Bruder} C. Bruder and H. Schoeller, Phys. Rev. Lett. {\bf
72}, 1076 (1994); C. Bruder and H. Schoeller, in {\em Quantum Dynamics
of Submicron Structures}, edited by H.~A. Cerdeira, B. Kramer, and
G. Sch\"on, NATO ASI Ser. E, Vol. 291 (Kluwer, Dordrecht, 1995),
p. 383.

\bibitem{Stoof} T.~H. Stoof and Y.~V. Nazarov, Phys. Rev. B {\bf 53},
1050 (1996).

\bibitem{staffordwingreen} C.~A. Stafford and N.~S. Wingreen,
Phys. Rev. Lett. {\bf 76}, 1916 (1996).

\bibitem{isawa} Y. Isawa, A. Okamoto, and T. Hatano, preprint.

\bibitem{BlochSiegert} F. Bloch and A. Siegert, Phys. Rev. {\bf 57},
522 (1940).

\bibitem{JaynCumm} E.~T. Jaynes and F.~W. Cummings, IEEE Proc. {\bf
51}, 89 (1963).

\bibitem{Shirley} J.~H. Shirley, Phys. Rev. {\bf 138}, B979 (1965).

\bibitem{HoltHone} M. Holthaus and D. Hone, Phys. Rev. B {\bf 47},
6499 (1993).

\end{references}
\end{document}